\documentstyle[aps,prd,twocolumn,epsf]{revtex}

\begin{document}
\draft
\title{Solutions of Conformal Gravity with Dynamical Mass Generation in the Solar
System}
\author{Joshua Wood and William Moreau}
\address{Department of Physics and Astronomy\\
University of Canterbury\\
Private Bag 4800\\
Christchurch, New Zealand}
\date{\today}
\maketitle

\begin{abstract}
The field equations of Mannheim's theory of conformal gravity with dynamic
mass generation are solved numerically in the interior and exterior regions
of a model spherically symmetric sun with matched boundary conditions at the
surface. The model consists of a generic fermion field inside the sun, and a
scalar Higgs field in both the interior and exterior regions. From the
conformal geodesic equations it is shown how an asymptotic gradient in the
Higgs field causes an anomalous radial acceleration in qualitative agreement
with that observed on the Pioneer 10/11, Galileo, and Ulysses spacecraft. At
the same time the standard solar system tests of general relativity are
preserved within the limits of observation.
\end{abstract}




\section{Introduction}

It has become increasingly probable in recent times that general relativity
is on its way to being falsified as a classical theory of gravitation.
General relativity fails to account for the motions of galactic clusters (
\cite{blumenthal84},\cite{summers95}) and rotations of individual galaxies (
\cite{rubin78}, \cite{ryden88}, see also \cite{christodoulou91} and
references therein) without the ad hoc assumption of varying amounts and
distributions of cold dark matter \cite{blumenthal84}. At the same time the
existence of cold dark matter in the universe is becoming more and more in
doubt. As determined by the Boomerang Experiment, the small amplitude of the
second peak relative to the first in the power spectrum of the cosmic
microwave background does not support the cold dark matter hypothesis \cite
{McGaugh-CMB00}. Even within the solar system general relativity has met
with a potential failure. Radio metric data from the Pioneer 10/11, Galileo,
and Ulysses spacecraft indicate a constant (with respect to radius from the
sun), radial acceleration directed toward the sun of magnitude $\sim
8.5\times 10^{-10}$ m s$^{-2}$, in addition to the expected
Newtonian-Einsteinian acceleration that varies as the inverse radius squared 
\cite{anderson1}. Furthermore, Viking Lander ranging data limit any such
anomalous radial accelerations on the planets Earth and Mars to less than
one hundredth of that detected on the spacecraft, a violation of the
equivalence principle. On a more fundamental plane, general relativity is
not compatible with dynamical mass generation in that the latter leads to a
manifestly traceless energy-momentum tensor as the source for Einstein's
field equation, while the Einstein tensor, the left side of the field
equation, is not traceless.

Collectively these problems pose a challenge to general relativity as a
classical theory of gravity, and it may be appropriate to seriously consider
alternatives. Two such alternatives have appeared in the literature in
recent years: (1) Milgrom's modified Newtonian dynamics (MOND) \cite
{milgrom83,sanders90}, and (2) the Weyl fourth-order, conformally invariant
gravitational theory which has been re-examined recently by Mannheim and
Kazanas \cite{mannhexsol} and extended by Mannheim to include dynamic mass
generation \cite{mannhdyn}. Both of these theories have been used to account
for galactic rotation curves without dark matter \cite{christodoulou91}, 
\cite{mannhgalrotcurves93,mannhgalrotcurves}. However, only (2) is
compatible with dynamical mass generation, in fact requires it, and
therefore we believe (2) is the stronger candidate as an alternative to
general relativity.

In this paper we investigate conformal gravity with dynamical mass
generation in the solar system to see whether the theory can account for the
anomalous, constant radial acceleration on the spacecraft while also passing
the standard solar system tests, namely light bending at the limb of the sun
and perihelion precession. As the field equations of conformal gravity are
nonlinear, fourth-order differential equations, exact analytic solutions are
not possible, even in a simple solar system model. We first consider
approximate solutions and then confirm the general validity of the
approximate results with complete numerical solutions of the interior and
exterior problems with matched boundary conditions at the limb of the sun.
We find that a completely conformal theory, including a scalar Higgs field
for dynamical mass generation, does, in fact, predict an anomalous,
nearly-constant radial acceleration toward the sun on a test particle such
as a spacecraft, while at the same time satisfying the standard solar system
tests. Furthermore, we give a qualitative argument why planets should behave
differently than spacecraft and not exhibit the same anomalous radial
acceleration.

\section{Conformal gravity with dynamic mass generation}

Conformal gravity is based on the Weyl geometry action, 
\begin{eqnarray}
I_{w} &=&-\alpha \int d^{4}x\sqrt{-g}C_{\mu \nu \lambda \rho }C^{\mu \nu
\lambda \rho }  \nonumber \\
&=&-2\alpha \int d^{4}x\sqrt{-g}\left( R_{\mu \nu }R^{\mu \nu }-\frac{1}{3}%
R^{2}\right) +\left( \text{surface term}\right) \,,  \label{weyl}
\end{eqnarray}
where $C_{\mu \nu \lambda \rho }$ is the Weyl conformal tensor, $g\equiv
\det \left( g_{\mu \nu }\right) $, $R_{\mu \nu \lambda \rho }$ is the
Riemann tensor, $R_{\mu \nu }\equiv R^{\lambda }{}_{\mu \lambda \nu }$ is
the Ricci tensor, $R\equiv R^{\mu }{}_{\mu }$ is the Ricci scalar, and $%
\alpha $ is a dimensionless constant. The Weyl action is invariant under a
conformal transformation, $g_{\mu \nu }\left( x\right) \rightarrow \left[
\Omega \left( x\right) \right] ^{2}g_{\mu \nu }\left( x\right) $, which is
an arbitrary, continuous deformation of the spacetime manifold. We use a
metric signature $\left( -,+,+,+\right) $ and the sign conventions of
Weinberg \cite{weinb1}.

For dynamic mass generation, following Mannheim \cite{mannhdyn}, we choose a
conformally invariant matter action in a model consisting of a
self-interacting scalar Higgs field $S$ and a generic fermion field $\psi $
with a Yukawa interaction with the scalar field as in the Standard Model of
particle physics. The requirement of conformal invariance restricts such a
matter action to the form 
\[
I_{m}=-\int d^{4}x\sqrt{-g}\left\{ i\bar{\psi}\left[ \gamma ^{\mu }\left(
x\right) \left( \partial _{\mu }+\Gamma _{\mu }\left( x\right) \right) +ihS%
\right] \psi \right. 
\]
\begin{equation}
\left. +\frac{1}{2}\partial _{\mu }S\partial ^{\mu }S-\frac{1}{2}\left( 
\frac{R}{6}\right) S^{2}+\lambda S^{4}\right\} \,,  \label{matter}
\end{equation}
where $\Gamma _{\mu }\left( x\right) $ is a spin connection, $h$ is a
dimensionless Yukawa coupling constant, and the quadratic term in $S$ with $%
R/6<0$ creates a situation of spontaneous symmetry breaking and consequent
dynamical mass generation.

Setting the variation of the total action, $I=I_{w}+I_{m}$, with respect to
the metric equal to zero yields the Weyl field equations, 
\begin{equation}
W_{\mu \nu }=\frac{1}{4\alpha }T_{\mu \nu }\,,  \label{feq}
\end{equation}
where the source, 
\begin{eqnarray}
T_{\mu \nu } &\equiv &\delta I_{m}/\delta g^{\mu \nu }=i\bar{\psi}\gamma
_{\mu }\left( x\right) \left[ \partial _{\nu }+\Gamma _{\nu }\left( x\right) %
\right] \psi  \nonumber \\
&&+\frac{2}{3}S_{\mu }S_{\nu }-\frac{1}{6}g_{\mu \nu }S_{\lambda }S^{\lambda
}-\frac{1}{3}SS_{\mu ;\nu }+\frac{1}{3}g_{\mu \nu }SS^{\lambda }{}_{;\lambda
}  \nonumber \\
&&-\frac{1}{6}S^{2}\left( R_{\mu \nu }-\frac{1}{2}g_{\mu \nu }R\right)
-g_{\mu \nu }\lambda S^{4}\,,  \label{emt}
\end{eqnarray}
is the energy-momentum tensor, which in this theory includes the scalar
Higgs field, and the tensor on the left side, 
\[
W_{\mu \nu }=-\frac{1}{6}g_{\mu \nu }R^{;\lambda }{}_{;\lambda }+\frac{2}{3}%
R_{;\mu ;\nu }+R_{\mu \nu }{}^{;\lambda }{}_{;\lambda }-R_{\mu }{}^{\lambda
}{}_{;\nu ;\lambda }-R_{\nu }{}^{\lambda }{}_{;\mu ;\lambda } 
\]
\begin{equation}
+\frac{2}{3}RR_{\mu \nu }-2R_{\mu }{}^{\lambda }R_{\lambda \nu }+\frac{1}{2}%
g_{\mu \nu }R_{\lambda \rho }R^{\lambda \rho }-\frac{1}{6}g_{\mu \nu }R^{2}
\label{bach}
\end{equation}
was first obtained by Bach \cite{bach21}.

Setting the variation of the action with respect to the two fields to zero
yields the two field equations of motion, 
\begin{equation}
\left\{ i\gamma ^{\mu }\left( x\right) \left[ \partial _{\mu }+\Gamma _{\mu
}\left( x\right) \right] -hS\right\} \psi =0\,,  \label{feqom}
\end{equation}
\begin{equation}
S^{\mu }{}_{;\mu }+\frac{R}{6}S-4\lambda S^{3}=-h\bar{\psi}\psi \,.
\label{heqom}
\end{equation}
Equation (\ref{feqom}) is the Dirac equation in curved spacetime with a
fermion mass due to spontaneous symmetry breaking given by 
\begin{equation}
m=hS.  \label{mass}
\end{equation}
Equation (\ref{heqom}) is the Klein-Gordon equation in curved spacetime with
a self interaction, $-4\lambda S^{3}$, and a fermion source, $-h\bar{\psi}%
\psi $. The scalar curvature, or more precisely $-R/6$, plays the role of
the mass squared of the Higgs boson. In contrast to the Standard Model of
particle physics, where the field point of the Higgs vacuum is a universal
constant over all spacetime, in the present context the Higgs field is
dynamically connected to the metric of spacetime through Eqs. (\ref{feq}), (%
\ref{feqom}), and (\ref{heqom}), and is not constant in general. In the
present problem the Higgs field varies radially from the sun, and it is the
radial gradient of the Higgs field that is responsible for the anomalous
radial acceleration.

The traceless property of the energy-momentum tensor with dynamic mass
generation is most easily established from Eq. (\ref{emt}) starting in a
gauge where the scalar Higgs field takes on a constant, nonzero value $S_{0}$%
. In this case, using the field equations of motion, Eqs. (\ref{feqom}) and (%
\ref{heqom}), we see that $T^{\mu }{}_{\mu }=0\,$. Since $T_{\mu \nu }$ is a
conformal tensor, under conformal transformation, $T^{\mu }{}_{\mu }=g^{\mu
\nu }T_{\mu \nu }\rightarrow \left[ \Omega ^{2}\left( x\right) \right]
^{-1}g^{\mu \nu }\Omega ^{-2}\left( x\right) T_{\mu \nu }=\Omega ^{-4}\left(
x\right) T^{\nu }{}_{\nu }\,$. In general the scalar Higgs field transform
as $S\left( x\right) \rightarrow S\left( x\right) /\Omega \left( x\right) $.
Therefore, in a general gauge in which the Higgs field is not constant, $%
S\left( x\right) =S_{0}/\Omega \left( x\right) $, and the trace of the
energy-momentum tensor is still zero since $T^{\nu }{}_{\nu }=\Omega
^{4}\left( x\right) T^{\mu }{}_{\mu }=0$.

The Bach tensor, the left side of the field equations, Eq. (\ref{feq}), is
also traceless, as can be seen from Eq. (\ref{bach}) which gives 
\begin{eqnarray*}
W^{\mu }{}_{\mu } &=&-\frac{2}{3}R^{;\lambda }{}_{;\lambda }+\frac{2}{3}%
R^{;\mu }{}_{;\mu }+R^{;\lambda }{}_{;\lambda }-2R^{\mu \lambda }{}_{;\mu
;\lambda } \\
&&+\frac{2}{3}R^{2}-2R^{\mu \lambda }R_{\mu \lambda }+2R_{\lambda \rho
}R^{\lambda \rho }-\frac{2}{3}R^{2} \\
&=&0\,,
\end{eqnarray*}
where we have used the covariant divergence of the contracted Bianchi
identity, $\left( R^{\mu \lambda }-g^{\mu \lambda }R/2\right) _{;\mu
;\lambda }=0$, to establish the cancellation of the third and fourth terms.
Thus dynamical mass generation is compatible with conformal gravity but not
with general relativity since the Einstein tensor is not traceless in
general: $E^{\mu }{}_{\mu }=R^{\mu }{}_{\mu }-\frac{1}{2}g^{\mu }{}_{\mu
}R=-R\neq 0$ .

In general relativity the predictive power of the theory lies in the
postulate that a test particle of mass $m$ follows a world line on the
spacetime manifold that minimizes the action \cite{ingsw-string1} 
\begin{equation}
I=-mc\int \,d\tau \,,  \label{geact}
\end{equation}
where $d\tau =\sqrt{-g_{\mu \nu }dx^{\mu }dx^{\nu }}$ is the line element.
With the mass $m$ of the test particle constant, the path is a geodesic.
However, the action of Eq. (\ref{geact}) is not suitable for a conformal
theory because it is not invariant under conformal transformation. In fact,
conformal gravity has been criticized by Perlick and Xu \cite{per&xu} as
being non-predictive except for null geodesics because the action of Eq. (%
\ref{geact}) is arbitrarily variable under conformal transformation. What
these authors have failed to recognize is that a conformal theory must be
completely conformal. The action of Eq. (\ref{geact}) must be replaced by a
conformally invariant action, and the way to make this change is apparent
from Eqs. (\ref{mass}) and (\ref{geact}). The mass of the test particle in a
conformal theory must no longer be a fixed attribute of the particle, but
instead it is a dynamic quantity generated through spontaneous broken
symmetry and given by Eq. (\ref{mass}). The mass of the test particle now
depends on position in spacetime and must be moved inside the integral, 
\begin{equation}
I=-c\int m\left( x\right) d\tau =-hc\int S\left( x\right) \,d\tau \,.
\label{conformact}
\end{equation}
The action of Eq. (\ref{conformact}) is invariant since under conformal
transformation $d\tau \rightarrow \Omega \left( x\right) d\tau $ and $%
S\left( x\right) \rightarrow \left[ \Omega \left( x\right) \right]
^{-1}S\left( x\right) $. Thus we see that a conformal theory of gravity is
not only compatible with dynamical mass generation but is non-predictive
without it.

Conformal geodesic equations are obtained from the conformal action given by
Eq. (\ref{conformact}) by requiring it to be stationary for arbitrary small
variations in path between two fixed points in spacetime: 
\[
-\frac{\delta S}{hc}=\delta \int_{P_{1}}^{P_{2}}S\left( x\right) \,d\tau
=\int_{P_{1}}^{P_{2}}\left[ \delta Sd\tau +S\delta \left( d\tau \right) %
\right] =0\,, 
\]
\[
\int_{P_{1}}^{P_{2}}\left\{ \frac{\partial S}{\partial x^{\sigma }}\delta
x^{\sigma }-\frac{S}{2}\left[ \frac{dx^{\mu }}{d\tau }\frac{dx^{\nu }}{d\tau 
}\frac{\partial g_{\mu \nu }}{\partial x^{\sigma }}\right. \right. 
\]
\[
\left. \left. +g_{\mu \sigma }\frac{dx^{\mu }}{d\tau }\frac{d\left( \delta
x^{\sigma }\right) }{d\tau }+g_{\sigma \nu }\frac{d\left( \delta x^{\sigma
}\right) }{d\tau }\frac{dx^{\nu }}{d\tau }\right] \right\} d\tau =0\,, 
\]
and, integrating by parts with $\delta x^{\sigma }\left( P_{1}\right)
=\delta x^{\sigma }\left( P_{2}\right) =0$, we have 
\[
\int_{P_{1}}^{P_{2}}\left\{ \frac{\partial S}{\partial x^{\sigma }}-\frac{1}{%
2}S\frac{dx^{\mu }}{d\tau }\frac{dx^{\nu }}{d\tau }\frac{\partial g_{\mu \nu
}}{\partial x^{\sigma }}\right. 
\]
\[
\left. +\frac{1}{2}\frac{d}{d\tau }\left[ S\left( g_{\mu \sigma }\frac{%
dx^{\mu }}{d\tau }+g_{\sigma \nu }\frac{dx^{\nu }}{d\tau }\right) \right]
\right\} \delta x^{\sigma }d\tau =0\,. 
\]
For arbitrary variations $\delta x^{\sigma }$ and multiplying by $g^{\rho
\sigma }/S$ and summing on $\sigma $, we obtain the conformal geodesic
equation, 
\begin{equation}
\frac{d^{2}x^{\rho }}{d\tau ^{2}}+\Gamma ^{\rho }{}_{\mu \nu }\frac{dx^{\mu }%
}{d\tau }\frac{dx^{\nu }}{d\tau }+\frac{1}{S}\frac{\partial S}{\partial
x^{\lambda }}\left( g^{\rho \lambda }+\frac{dx^{\rho }}{d\tau }\frac{%
dx^{\lambda }}{d\tau }\right) =0\,.  \label{geodesic}
\end{equation}

In their original paper on conformal gravity Mannheim and Kazanas \cite
{mannhexsol} obtained an analytic solution of the homogeneous equations, 
\begin{equation}
W_{\mu \nu }=0\,,  \label{falsevac}
\end{equation}
for a static, spherically-symmetric metric corresponding to the line
element, 
\begin{equation}
ds^{2}=-b\left( r\right) c^{2}dt^{2}+\frac{1}{b\left( r\right) }%
dr^{2}+r^{2}\left( d\theta ^{2}+\sin ^{2}\theta d\phi ^{2}\right) \,.
\label{conformle}
\end{equation}
It is given by 
\begin{equation}
b\left( r\right) =1-\frac{\beta \left( 2-3\beta \gamma \right) }{r}-3\beta
\gamma +\gamma r-\kappa r^{2}\,,  \label{kmsol}
\end{equation}
where $\beta $, $\gamma $, and $\kappa $ are integration constants.

As Mannheim and Kazanas have shown, the above simple form of the metric
suffices in a conformal theory. Starting with the general line element in
terms of a radial coordinate $\rho $, 
\begin{equation}
ds^{2}=-B\left( \rho \right) c^{2}dt^{2}+A\left( \rho \right)
dr^{2}+r^{2}\left( d\theta ^{2}+\sin ^{2}\theta d\phi ^{2}\right) \,,
\label{linelmnt}
\end{equation}
one makes a change of radial variable $\rho =p\left( r\right) $ and rewrites
the line element as 
\begin{eqnarray}
ds^{2} &=&\left[ \frac{p\left( r\right) }{r}\right] ^{2}\left[ -b\left(
r\right) c^{2}dt^{2}+a\left( r\right) dr^{2}\right.  \nonumber \\
&&\left. +r^{2}\left( d\theta ^{2}+\sin ^{2}\theta d\phi ^{2}\right) \right]
\,\,,  \label{form2}
\end{eqnarray}
where $b\left( r\right) \equiv r^{2}B\left( p\left( r\right) \right) /\left[
p\left( r\right) \right] ^{2}$ and $a\left( r\right) \equiv r^{2}A\left(
p\left( r\right) \right) \left( dp/dr\right) ^{2}/\left[ p\left( r\right) %
\right] ^{2}$. Then, if one chooses the function $p\left( r\right) $ to
satisfy 
\begin{equation}
\frac{1}{p\left( r\right) }=-\int \frac{dr}{r^{2}\sqrt{A\left( p\left(
r\right) \right) B\left( p\left( r\right) \right) }}\,,  \label{pr}
\end{equation}
the line element becomes 
\begin{eqnarray}
ds^{2} &=&\left[ \frac{p\left( r\right) }{r}\right] ^{2}\left[ -b\left(
r\right) c^{2}dt^{2}+\frac{1}{b\left( r\right) }dr^{2}\right.  \nonumber \\
&&\left. +r^{2}\left( d\theta ^{2}+\sin ^{2}\theta d\phi ^{2}\right) \right]
\,,  \label{oneconaway}
\end{eqnarray}
which is related to the simple line element of Eq. (\ref{conformle}) by a
conformal transformation with $\Omega \left( r\right) =r/p\left( r\right) $.

Mannheim and Kazanas \cite{mannhexsol} interpret Eq. (\ref{kmsol}) as an
exterior vacuum solution with a Schwarzschild-like metric in a background de
Sitter spacetime, where the integration constant $\gamma $ measures the
departure from the Schwarzschild metric. They have argued that with $\gamma
\ll 1$ the metric of Eq. (\ref{kmsol}) enjoys all of the successes of
general relativity in the solar system, while on larger interstellar scales
in a galaxy the linear term $\gamma r$ can provide the increasing
gravitational potential with radius required to supplement the decreasing
Newtonian potential in order to account for the plateau characteristics of
galactic rotation curves \cite{mannhgalrotcurves}. They have also argued
that the integration constant $\kappa $ is so small that the de Sitter term
is negligible except on cosmological distance scales. Perlick and Xu \cite
{per&xu} correctly have insisted that the values of the integration
constants, $\beta ,\gamma ,$ and $\kappa $, cannot be chosen at will, but
must be determined by matching boundary conditions with an interior solution
at the inner boundary with the source and at infinity.

But there is an even more serious criticism of the vacuum solution of
Mannheim and Kazanas, and that is it is not a vacuum solution at all. The
vacuum in conformal gravity is not $T_{\mu \nu }=0$ because even outside the
source the Higgs field must be nonzero. Otherwise a massive test particle
would not be possible. Even in a gauge where the Higgs field is constant,
outside the source the final term in Eq. (\ref{emt}) contributes to a
nonzero energy-momentum tensor.\ The next to last term also contributes for
all metrics in which the Ricci tensor is nonzero, including the metric of
Mannheim and Kazanas given by Eq. (\ref{kmsol}). In a general gauge all the
terms in Eq. (\ref{emt}) except the first contribute outside the source. We
have argued above that conformal gravity not only is compatible with
dynamical mass generation through spontaneous symmetry breaking of a Higgs
vacuum, but also requires it in order to be predictive. It is therefore not
logically consistent to ignore the Higgs contribution to the energy-momentum
tensor outside the source, and the vacuum solution of Mannheim and Kazanas
is not physically relevant by itself. All we can say is that mathematically
it is an exact complimentary solution of the homogeneous field equations
that may be a good approximation to an exterior particular solution.

\section{Approximate analytic considerations}

Before presenting our matched interior-exterior numerical solutions we begin
our investigation by asking the question, what are the characteristics that
would be required of the Higgs field in order to produce an anomalous,
constant radial acceleration at large radii in the solar system such as that
observed on Pioneer 10/11 etc. spacecraft, while at the same time permitting
the theory to pass the standard tests in the inner solar system. This
question can be answered by the conformal geodesic equation. We wish
initially to answer it analytically, so we have to make a few simplifying
assumptions. First, since the theory must pass the standard solar system
tests, the metric must be very close to the Schwarzschild metric, and we
will assume initially that it is exactly Schwarzschild. One way to assess
this assumption is that the Schwarzschild metric satisfies $R_{\mu \nu }=0$,
and therefore it also satisfies $W_{\mu \nu }=0$. So the Schwarzschild
metric is a first approximation to the conformal metric with $T_{\mu \nu
}\approx 0$. As we see from Eq. (\ref{emt}), the energy momentum tensor will
be small outside the sun if the radial gradient of the Higgs field $S$ and
the self coupling constant $\lambda $ are both small with respect to unity.
We will see later from our numerical calculations that both of these
assumptions are valid in the asymptotic region. Note that for the
Schwarzschild metric the next to last term in $T_{\mu \nu }$ is identically
zero since the scalar curvature $R=R^{\mu }{}_{\mu }=0$. But realistically
the metric must be near to Schwarzschild, but not exactly Schwarzschild
because the mass squared of the Higgs particle is $R/6$.

Assuming spherical symmetry such that $S=S\left( r\right) $ and for a line
element of the form of Eq. (\ref{conformle}) with 
\begin{equation}
b\left( r\right) =1-\frac{2m}{r}\,  \label{schw}
\end{equation}
where $m\equiv GM/c^{2}$ is the geometric radius of the sun, the conformal
geodesic equation, Eq. (\ref{geodesic}), evaluated for $\rho =0,1,2,3$ is
given respectively by $\left( c=1\right) $ 
\begin{equation}
\frac{d}{d\tau }\left( bS\frac{dt}{d\tau }\right) =0\,,  \label{geo1}
\end{equation}
\[
\frac{d^{2}r}{d\tau ^{2}}+\frac{1}{2}b\frac{db}{dr}\left( \frac{dt}{d\tau }%
\right) ^{2}-\frac{1}{2b}\frac{db}{dr}\left( \frac{dr}{d\tau }\right)
^{2}-br\left( \frac{d\theta }{d\tau }\right) ^{2} 
\]
\begin{equation}
-br\sin ^{2}\theta \left( \frac{d\phi }{d\tau }\right) ^{2}+\frac{1}{S}\frac{%
dS}{dr}\left[ b+\left( \frac{dr}{d\tau }\right) ^{2}\right] =0\,,
\label{geo2}
\end{equation}
\begin{equation}
\frac{d^{2}\theta }{d\tau ^{2}}+\frac{2}{r}\frac{dr}{d\tau }\frac{d\theta }{%
d\tau }-\sin \theta \cos \theta \left( \frac{d\phi }{d\tau }\right) ^{2}+%
\frac{1}{S}\frac{dS}{dr}\frac{d\theta }{dr}\frac{dr}{d\tau }=0\,,
\label{geo3}
\end{equation}
\begin{equation}
\frac{d^{2}\phi }{d\tau ^{2}}+2\cot \theta \frac{d\theta }{d\tau }\frac{%
d\phi }{d\tau }+\frac{2}{r}\frac{dr}{d\tau }\frac{d\phi }{d\tau }+\frac{1}{S}%
\frac{dS}{dr}\frac{d\phi }{dr}\frac{dr}{d\tau }=0\,.  \label{geo4}
\end{equation}

For spherical symmetry, we lose no generality by restricting motion to the $%
\theta =\pi /2$ plane, in which case Eq. (\ref{geo3}) is trivially
satisfied. Equations (\ref{geo1}) and (\ref{geo4}) may be integrated once ($%
l $ and $k$ are constants of integration), 
\begin{equation}
\frac{dt}{d\tau }=\frac{l}{bS}\,,  \label{geo1int}
\end{equation}
\begin{equation}
r^{2}\frac{d\phi }{d\tau }=\frac{k}{S}\,,  \label{geo4int}
\end{equation}
and the result substituted into Eq. (\ref{geo2}) giving 
\[
\frac{d^{2}r}{d\tau ^{2}}+\frac{1}{2b}\frac{db}{dr}\left[ \frac{l^{2}}{S^{2}}%
-\left( \frac{dr}{d\tau }\right) ^{2}\right] 
\]
\begin{equation}
+\frac{1}{S}\frac{dS}{dr}\left[ b+\left( \frac{dr}{d\tau }\right) ^{2}\right]
-\frac{k^{2}b}{r^{3}S^{2}}=0\,.  \label{geo2sub}
\end{equation}
From the line element with $d\theta =0$ and $l/bS$ substituted for $dr/d\tau 
$ according to Eq. (\ref{geo1int}), we write 
\begin{equation}
1=\frac{l^{2}}{bS^{2}}-\frac{1}{b}\left( \frac{dr}{d\tau }\right) ^{2}-\frac{%
k^{2}}{r^{2}S^{2}}\,.  \label{lemod}
\end{equation}
Solving Eq. (\ref{lemod}) for $\left( dr/d\tau \right) ^{2}$ and
substituting into Eq. (\ref{geo2sub}) gives 
\begin{equation}
\frac{d^{2}r}{d\tau ^{2}}+\frac{1}{2}\frac{db}{dr}\left( 1+\frac{k^{2}}{%
r^{2}S^{2}}\right) +\frac{1}{S^{3}}\frac{dS}{dr}\left( l^{2}-k^{2}b\right) -%
\frac{k^{2}b}{r^{3}S^{2}}=0\,,  \label{radeqgen}
\end{equation}
and finally for radial motion $\left( k=0\right) $, 
\begin{equation}
\frac{d^{2}r}{d\tau ^{2}}+\frac{1}{2}\frac{db}{dr}+\frac{l^{2}}{S^{3}}\frac{%
dS}{dr}=0\,.  \label{radeq}
\end{equation}

As the radiometric measurements of the accelerations of the spacecraft in
the solar system are based on coordinate rather than proper time, we convert
Eq. (\ref{radeq}) to coordinate time using $dt/d\tau =l/bS$ and restore $c$
by letting $t\rightarrow ct$, 
\[
\frac{d^{2}r}{dt^{2}}-\left( \frac{1}{b}\frac{db}{dr}+\frac{1}{S}\frac{dS}{dr%
}\right) \left( \frac{dr}{dt}\right) ^{2} 
\]
\begin{equation}
+\frac{b^{2}c^{2}S^{2}}{2l^{2}}\frac{db}{dr}+\frac{b^{2}c^{2}}{S}\frac{dS}{dr%
}=0\,.  \label{radeqct}
\end{equation}
It is the final term in Eq. (\ref{radeqct}) that has the potential of
accounting for a constant radial acceleration for large radius. In this
limit $b=1-2m/r\rightarrow 1$, so we set 
\begin{equation}
c^{2}\frac{d\ln S}{dr}=a\left( r\right) \,,  \label{acconst}
\end{equation}
where $a\left( r\right) $ is an anomalous acceleration. The next to final
term in Eq. (\ref{radeqct}) accounts for the Newtonian acceleration. 
\begin{equation}
\frac{b^{2}c^{2}S^{2}}{2l^{2}}\frac{db}{dr}=\frac{1}{l^{2}}\left( 1-\frac{2m%
}{r}\right) ^{2}S^{2}\frac{GM}{r^{2}}\rightarrow \frac{S^{2}}{l^{2}}\frac{GM%
}{r^{2}}  \label{na}
\end{equation}
since for $r=40$ AU, $2m/r=2.0\times 10^{-8}\ll 1$. Turning our attention to
the second term in Eq. (\ref{radeqct}), we see from Eqs. (\ref{schw}) and (%
\ref{acconst}) that 
\[
\left( \frac{1}{b}\frac{db}{dr}+\frac{1}{S}\frac{dS}{dr}\right) \left( \frac{%
dr}{dt}\right) ^{2}=\left[ \frac{\frac{2m}{r^{2}}}{1-\frac{2m}{r}}+\frac{%
a\left( r\right) }{c^{2}}\right] \left( \frac{dr}{dt}\right) ^{2} 
\]
\begin{equation}
\simeq \left( \frac{2GM}{r^{2}}+a\left( r\right) \right) \left( \frac{1}{c}%
\frac{dr}{dt}\right) ^{2}\,,  \label{st}
\end{equation}
which is negligible for nonrelativistic speeds.

Based upon the above stated assumptions and approximations, the conformal
geodesic equation, for nonrelativistic radial motion with a Schwarzschild
metric, reduces asymptotically to 
\begin{equation}
\frac{d^{2}r}{dt^{2}}=-\frac{S^{2}}{l^{2}}\frac{GM}{r^{2}}-a\left( r\right)
\,.  \label{phiques}
\end{equation}
The requirement, based upon the radiometric data from the spacecraft, that $%
a\left( r\right) =a_{0}$ be asymptotically constant determines the
asymptotic form that the scalar Higgs field must take through Eq. (\ref
{acconst}): 
\begin{equation}
\frac{d\ln S}{dr}=\frac{a_{0}}{c^{2}}\,,\quad \ln S=\frac{a_{0}}{c^{2}}r+\ln
S_{0}\,,\quad S=S_{0}\exp \left( \frac{a_{0}}{c^{2}}r\right) \,.
\label{asymphiggs}
\end{equation}
For $a_{0}=8.5\times 10^{-10}$ m s$^{-2}$ and $r=40$ AU, $%
a_{0}r/c^{2}=5.6\times 10^{-14}\ll 1$ and the required asymptotic form is
well approximated by 
\begin{equation}
S=S_{0}\left( 1+\frac{a_{0}}{c^{2}}r\right) \,.  \label{asympform}
\end{equation}
Then the Newtonian term in Eq. (\ref{phiques}) becomes asymptotically 
\begin{equation}
-\frac{S^{2}}{l^{2}}\frac{GM}{r^{2}}=-\frac{\left( S_{0}\right) ^{2}}{l^{2}}%
\left( 1+\frac{a_{0}}{c^{2}}r\right) ^{2}\frac{GM}{r^{2}}\simeq -\frac{GM}{%
r^{2}}\,,  \label{newt}
\end{equation}
if we set the two integration constants equal, $l=S_{0}$, and neglect $%
a_{0}r/c^{2}$ with respect to $1$. Note from Eq. (\ref{geo1int}) that $l$
sets the scale of proper time, and we are free to measure proper time in any
units we wish. Finally, if the Higgs field takes on the asymptotic form of
Eq. (\ref{asymphiggs}), then the conformal geodesic equation becomes in the
asymptotic region 
\begin{equation}
\frac{d^{2}r}{dt^{2}}=-\frac{GM}{r^{2}}-a_{0}\,.  \label{final}
\end{equation}

While Eq. (\ref{asympform}) gives the required asymptotic form of the Higgs
field in order for the conformal theory to account for the anomalous radial
acceleration, it remains to be determined whether the conformal field
equations have such an asymptotic solution. The answer to this question
requires the full interior and exterior solutions with matched boundary
conditions at the limb of the sun. But before turning our attention to a
full solution of the field equations, it is instructive to consider an
approximate analytic solution of Eq. (\ref{heqom}), the equation of motion
of the Higgs field, in the exterior region $\left( \psi =0\right) $,
assuming the metric is exactly Schwarzschild $\left( R=0\right) $ and
neglecting the self-interaction term $\left( \lambda =0\right) $. In this
case and with spherical symmetry Eq. (\ref{heqom}) becomes 
\begin{equation}
S^{;\mu }{}_{;\mu }=\frac{d^{2}S}{dr^{2}}+\frac{dS}{dr}\left( \frac{2m}{%
r^{2}-2mr}+\frac{2}{r}\right) =0,  \label{heqomapprox}
\end{equation}
and this differential equation has a solution, 
\begin{equation}
S=C_{1}\ln \left( 1-\frac{2m}{r}\right) +C_{2}\,.  \label{analyhiggssol}
\end{equation}
Although there are two constants of integration in the solution of the
second-order differential equation, physically only their ratio is
important, as can be seen from the conformal geodesic equation, Eq. (\ref
{geodesic}), in which for the present static case with spherical symmetry
the Higgs dependent factor in Eq. (\ref{geo2sub}) is 
\begin{equation}
\frac{1}{S}\frac{dS}{dr}=\left( \frac{2m}{r^{2}}\right) \left( 1-\frac{2m}{r}%
\right) ^{-1}\left[ \ln \left( 1-\frac{2m}{r}\right) +\frac{C_{2}}{C_{1}}%
\right] ^{-1}\,.  \label{physical}
\end{equation}
Thus we can choose either $C_{1}$ or $C_{2}$ arbitrarily and it is
convenient to evaluate $C_{2}$ by setting $S=S_{0}=1$ m$^{-1}$ at $r=R$ at
the limb of the sun. Then we have 
\begin{equation}
C_{2}=1-C_{1}\ln \left( 1-\frac{2m}{R}\right) \,.  \label{limber}
\end{equation}
Ideally we should evaluate $C_{1}$ by matching the slope, 
\begin{equation}
\frac{dS }{dr}=\frac{2mC_{1}}{r^{2}-2mr}\,,  \label{slope}
\end{equation}
with the slope of an interior solution at the limb of the sun. But since at
the present stage of the exposition we do not have an interior solution, we
instead evaluate $C_{1}$ by the observational requirement of an anomalous
radial acceleration of $a_{0}=-8.5\times 10^{-10}$ ms$^{-2}$ at a radius of $%
r=R_{40}=$ $40$ AU, giving 
\[
C_{1}=S_{0}\left[ \ln \left( 1-\frac{2m}{R_{40}}\right) -\ln \left( 1-\frac{%
2m}{R}\right) \right. 
\]
\begin{equation}
\left. -\frac{2mc^{2}}{a_{0}R_{40}^{2}}\left( 1-\frac{2m}{R_{40}}\right)
^{-1}\right] ^{-1}.  \label{see1}
\end{equation}

\begin{figure}[tbp]
\epsfxsize=13cm
\epsfbox{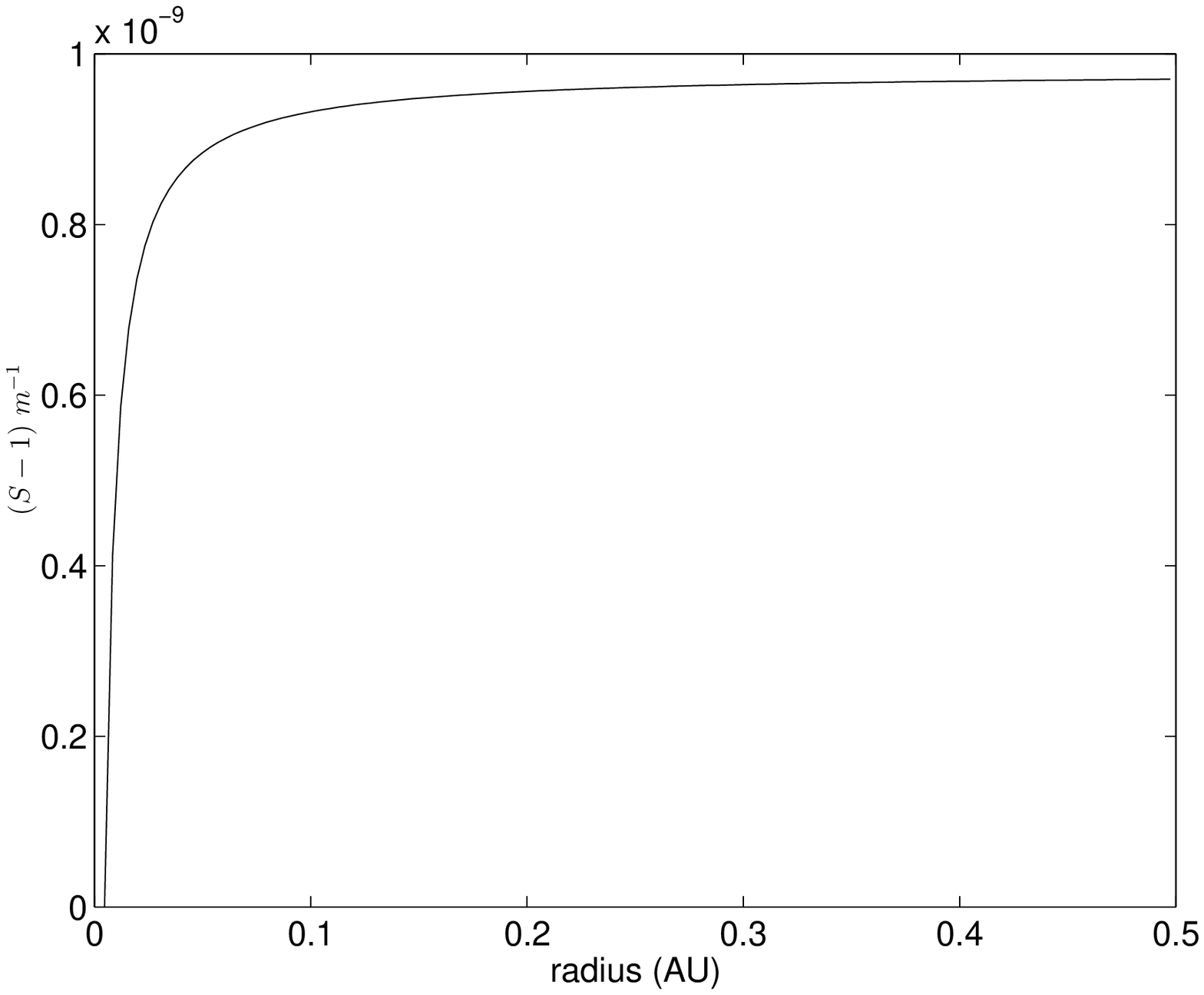}
\caption{Exterior Scalar Field}
\label{fesfield}
\end{figure}

Evaluating Eqs. (\ref{see1}) and (\ref{limber}), we obtain $C_{1}=1.20\times
10^{-4}$ m$^{-1}$ and $C_{2}=1+5.14\times 10^{-10}\approx 1$ m$^{-1}$. With
these constants, the Higgs field $S$ is found to have a value that is nearly
constant for $r>0.1$ AU, with a slope that is positive and very small with
respect to unity (see Fig. \ref{fesfield}, noting that only the first half
AU is shown for clarity, which establishes the asymptotic trend). This
approximate solution of the scalar field equation of motion shows that the
features the scalar field can be expected to have are intuitively
reasonable. Although the slope is not constant as would be required by
observation, the inclusion of a nearly, but not exactly, Schwarzschild
metric and a small but nonzero self coupling is enough to reduce the
variation with radius significantly, as we shall see. The approximations
used (zero Higgs mass and self coupling), although quite sweeping, are not
so extreme as to take the analytic solution of the reduced equation far from
the solutions of the full equation.

\section{Solar system tests}

Two further questions need to be considered, and these are the effect of the
scalar field on the deflection of light and perihelion precession, the
standard solar system tests of any gravitational theory. These questions can
be answered by the orbit equation ($u\equiv 1/r$ and $h$ is the angular
momentum per unit mass), 
\begin{equation}
\frac{d^{2}u}{d\phi ^{2}}=-\frac{db}{du}\frac{u^{2}}{2}-bu-\frac{1}{h^{2}}%
\left( \frac{S^{2}}{2}\frac{db}{du}+bS\frac{dS}{du}\right) \,,  \label{orbeq}
\end{equation}
which is derived from the conformal geodesic equation, Eq. (\ref{geodesic}),
for the case of the conformal line element given by Eq. (\ref{conformle}).
Regarding the deflection of light, we note that the angular momentum per
unit mass of a photon is infinite, and therefore the final term in Eq. (\ref
{orbeq}) involving the scalar field is zero. This result is compatible with
the fact that the photon, being massless, does not couple directly with the
Higgs field. Thus there is no direct effect on the deflection of light.
Indirectly the prediction of the conformal theory will differ from that of
general relativity by the degree that $b\left( r\right) $ differs from the
Schwarzschild metric. We will see in the numerical solution that this
difference is minimal. On the question of perihelion precession, while the
final term in Eq. (\ref{orbeq}) is nonzero we note that it does not contain
the inverse radius, $u$. In the standard iterative solution of Eq. (\ref
{orbeq}) assuming small eccentricity, the zeroth order approximation, $%
u=u_{0}\left( 1+e\cos \phi \right) $, is substituted into the right-hand
side of the equation, resulting in a secular term, $\left(
3m^{3}/h^{4}\right) e\phi \sin \phi $, that grows with the angle traversed.
Again, the presence of the Higgs field does not contribute directly to the
secular term and consequently to perihelion precession. Only through the
deviation of $b\left( r\right) $ from Schwarzschild will the prediction of
the conformal theory differ from general relativity.

\section{Numerical Solution}

Finally we present a numerical solution of the Weyl field equations and the
scalar field equation of motion that has been carried out over both the
interior and exterior domains, with boundary condition matching at the limb
of the source, a prototypical sun which is presumed static and spherically
symmetric. The exterior domain of the numerical solution has been extended
in fine resolution to a radius of 60 AU, and in coarser resolution to
further radius. The simple solar model used to perform these solutions was a
polytrope of order three, and the equations were solved for the metric
coefficient $b$ and the scalar field $S$, with the metric coefficient $a$
removed by coordinate and conformal transformations. The fermion field was
handled by following the averaging procedure used in \cite{mannhdyn} and
solving in terms of generic fermion number density and pressure. The main
methods of solution were Runge-Kutta integrations and Newton-Raphson
relaxations.

Boundary conditions at the limb of the source were provided by requiring
compatibility with observation. For the metric field, the Schwarzschild
solution is known to be very accurate in the inner solar system for light
deflection and precession observations, so the metric field was required to
match well with the Schwarzschild solution at the limb of the sun. The
scalar field has it's value normalised to one at this point. The limb
gradient of the scalar field can be limited in magnitude by compatibility
with existing observations, or a more specific value can be produced by
reference to the anomalous motion of the Pioneer spacecraft. If the observed
acceleration of these spacecraft is accurate (and a transponder signal
should be the most sensitive test we have of these effects to date), the
scalar field gradient corresponding to the observed acceleration at 40 AU
would be $S^{\prime }\sim 10^{-24}$ m$^{-2}$. The exterior numerical
solution gives a corresponding limb value of $S^{\prime }\sim 10^{-19}$ m$%
^{-2}$.

\begin{figure}[tbp]
\epsfxsize=13cm
\epsfbox{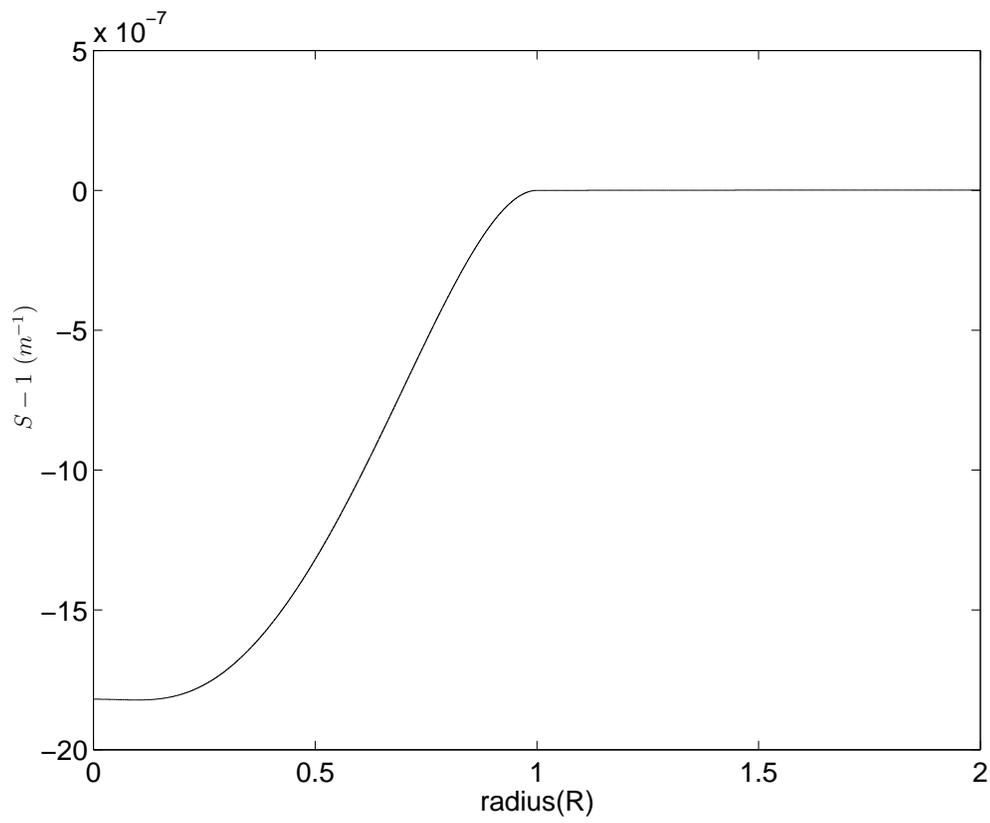}
\caption{Scalar Field - Boundary Matching}
\label{fsfield}
\end{figure}

\begin{figure}[tbp]
\epsfxsize=13cm
\epsfbox{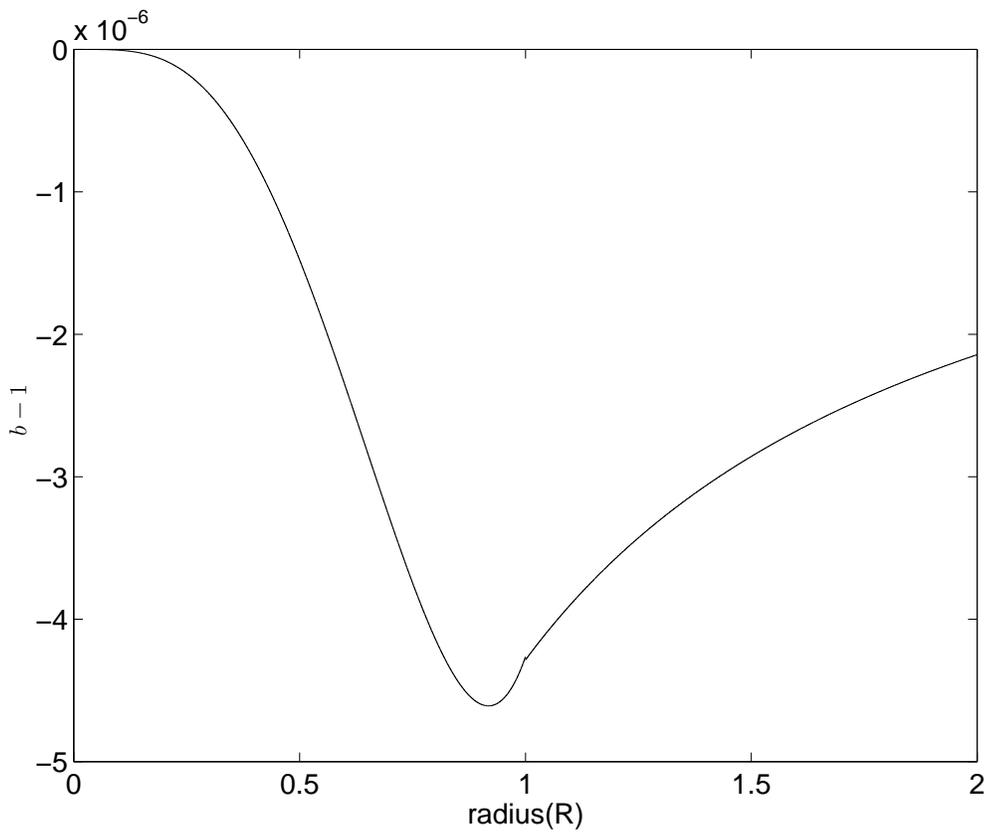}
\caption{Metric Coefficient - Boundary Matching}
\label{fbfield}
\end{figure}

With these boundary conditions in hand, a consistent interior/exterior
numerical solution was produced (Figures \ref{fsfield}, \ref{fbfield}). The
interior metric field merges well into an exterior solution of the
Schwarzschild form. The scalar field increases across the interior domain
before smoothly assuming a very slight gradient as required in the exterior
region. Newton-Raphson iterations on the exterior solution near the source
show that the metric field produced is in agreement with Schwarzschild to
one part in $10^{15}$, even in the presence of a scalar field. From our
arguments given above, we would conclude that theory passes the standard
solar system tests to this accuracy.

The far-field (30 -- 60 AU) behaviour of the fields is also satisfactory.
The metric field continues to agree with Schwarzschild to good order, up to
the accuracy limits of the numerical methods. The scalar field assumes a
very slight and asymptotically decreasing gradient which varies very slowly
across the exterior domain. The results are very similar to the restricted
analytic solution described above, but with the inclusion of the metric
field in the mass term resulting in a slower variation of the scalar field
gradient. It is expected that with improving accuracy of the far-field
methods that the scalar field will vary more slowly again.

The values that the constants of the theory assumed with these boundary
conditions were $\alpha \sim 3.5\times 10^{15}$ and $\lambda \leq 10^{-45}$.
The value of $\alpha $ is set through the boundary conditions, and that of $%
\lambda $ by the stability of the solution at large radius, larger values
producing runaway behaviour in the fields at large radius due to the $%
\lambda S^{3}$ term in the equations.

\section{Planetary Motion}

The form we have found for the scalar field shows a radially and rapidly
decreasing gradient in the vicinity of the source. This feature could have
relevance as to why anomalous accelerations are not observed for planetary
bodies, at least to the accuracy of ranging experiments to date (\cite
{anderson1}; see also comments in \cite{pre-guruprasad}). While it is not
yet known how the scalar field sourced by more than one body would be
configured, we expect that a secondary body, such as a planet, would
superpose a similar characteristic well upon the background field produced
by the primary. Such a superposition may well result in an average of the
scalar field gradient over the volume of the planet that is less than the
single-source field gradient. Thus the constituent particles of the planet
could, on average, experience less of an anomalous radial acceleration
towards the sun than they would if the rest of the planet were not present.
In this respect a planet is unlike a spacecraft, which would not appreciably
affect the scalar field in its own vicinity.

\section{Conclusion}

In this paper we have investigated the theory of conformal gravity with
dynamical mass generation, as originally presented by Mannheim \cite
{mannhdyn}. The theory has been presented here in a logically consistent
format that clarifies some areas in the construction of the theory that have
been confused in the past. In particular it has been emphasized here that
since conformal gravity not only is compatible with dynamical mass
generation, but also requires it to be predictive, the scalar Higgs field is
an integral part of the theory and cannot be ignored. From this framework,
analytic and numerical results have been extracted. The import of these
results is that it appears that conformal gravity with dynamical mass
generation can reproduce the gravitational effects within the solar system,
including the possible Pioneer spacecraft accelerations, while still
satisfying the standard solar system tests. This successful outcome was made
possible through the inclusion of the scalar field required for dynamical
mass generation. Solutions of the full system of equations have not been
obtained before.





%
%
\clearpage

\clearpage

\clearpage

\clearpage

\clearpage

\clearpage

\end{document}